\documentclass[5p]{elsarticle}

\usepackage{lineno,hyperref}
\modulolinenumbers[5]
\journal{Journal of Magnetism and Magnetic Materials}

\usepackage{color}
\usepackage{amsmath}
\usepackage{amssymb}
\usepackage{graphicx}
\usepackage{siunitx}
\usepackage{subfigure}
\begin{document}
\begin{frontmatter}
\author[label1]{M. Fries\corref{cor1}}
\ead{fries@fm.tu-darmstadt.de}
\author[label2]{T. Gottschall}
\author[label1,label3]{F. Scheibel}
\author[label1]{L. Pfeuffer}
\author[label1]{K. P. Skokov}
\author[label2]{I. Skourski}
\author[label3]{M. Acet}
\author[label3]{M. Farle}
\author[label2,label4]{J. Wosnitza}
\author[label1]{O. Gutfleisch}
\cortext[cor1]{Corresponding author}
\address[label1]{Institut f\"ur Materialwissenschaft, Technische Universit\"at Darmstadt, 64287 Darmstadt, Germany}
\address[label2]{Dresden High Magnetic Field Laboratory (HLD-EMFL), Helmholtz-Zentrum Dresden-Rossendorf, 01328 Dresden, Germany}
\address[label3]{Fakult\"at f\"ur Physik und CENIDE, Universit\"at Duisburg-Essen, 47057 Duisburg, Germany}
\address[label4]{Institut f\"ur Festk\"orper- und Materialphysik, Technische Universit\"at Dresden, 01062 Dresden, Germany}

\title{Dynamics of the magnetoelastic phase transition and adiabatic temperature change in Mn$_{1.3}$Fe$_{0.7}$P$_{0.5}$Si$_{0.55}$}

\begin{abstract}
The adiabatic temperature change $\Delta T_{ad}$ of a Mn$_{1.3}$Fe$_{0.7}$P$_{0.5}$Si$_{0.55}$ Fe$_2$P-type alloy was measured under different magnetic field-sweep rates from \SI{0.93}{Ts^{-1}} to \SI{2870}{Ts^{-1}}. We find a field-sweep-rate independent magnetocaloric effect due a partial alignment of magnetic moments in the paramagnetic region overlapping with the magnetocaloric effect of the first-order phase transition. Additionally, the first-order phase transition is not completed even in fields up to \SI{20}{T} leading to a non-saturating behavior of $\Delta T_{ad}$. Measurements in different pulsed fields reveal that the first-order phase transition cannot follow the fast field changes as previously assumed, resulting in a distinct field-dependent hysteresis in $\Delta T_{ad}$.
\end{abstract}


\end{frontmatter}

\section{Introduction}

The world is in constant need of easily available and environmentally friendly cooling technologies because of a predicted exponential worldwide increase in the demand of cooling power during the next 50 years~\cite{Morna2009}. A solid-state technology based on the known caloric effects \cite{Faehler2012, Gutfleisch2011, Moya2014, Gottschall2018a} is predicted to be more environmentally friendly and more energy efficient compared to conventional compressor-based refrigerators. Among those, magnetocaloric cooling is the most advanced of the different caloric concepts \cite{Kitanovski2014-book}. It is based on the magnetocaloric effect (MCE) manifesting itself in the temperature change of a magnetic material that is magnetized or demagnetized \cite{Weiss1917}. In a cyclic way, this temperature change can be used for cooling, in analogy to the gas-vapor compression cycle \cite{Brown1976}. 

In order to make magnetic refrigeration commercially viable, efficient materials need to be designed which can sustain high magnetic-field cycling frequencies~\cite{Kitanovski2014-book, Kuzmin2007, Smith2012}. 
The adiabatic temperature change $\Delta T_{ad}$ and the transitional entropy change $\Delta S_t$ both have to be maximized~\cite{Liu2012, Gutfleisch2016, Scheibel2018a}. Materials with a first-order phase transition are especially promising as they can show a large MCE under moderate magnetic-field changes~\cite{Franco2012}. Among these, Fe$_2$P-based compounds show a large MCE \cite{Yan2006, Yibole2014, Fries2017} and have a large potential for commercial application \cite{Gauss2016}. For Boron doped compounds useful secondary engineering properties such as mechanical integrity and high thermal conductivity are reported~\cite{Guillou2014-2, Guillou2014-3}. The material undergoes a hexagonal, isostructural phase transformation with only a small volume change of the unit cell~\cite{Dung2011-2}. Previous measurements of the MCE performed on a Mn-Fe-P-As (Fe$_2$P-type) compound in a \SI{1}{T} field change suggest that the material can follow field-cycling frequencies up to \SI{150}{Hz} \cite{Cugini2016}. However, a direct confirmation is still missing. It was shown previously for Heusler and antiperovskite materials that non-trivial field-rate dependences are observed for first-order phase transformations \cite{Gottschall2016-2, Zavareh2015, Scheibel2015}. A detailed investigation of the sweep-rate dependence of the adiabatic temperature change of Fe$_2$P-based alloys, one of the commercially most viable magnetocaloric material, is lacking so far.

\section{Experimental Details}

Polycrystalline Mn$_{1.3}$Fe$_{0.7}$P$_{0.5}$Si$_{0.55}$ was prepared by a powder-metallurgical process as described in \cite{Dung2011-2, Fries2017}. Magnetization measurements were performed using a commercial Quantum Design PPMS-14T VSM magnetometer with a heating and cooling rate of \SI{2}{K min^{-1}}. Direct measurements of the adiabatic temperature change $\Delta T_{ad}$ were carried out in two different setups. The ``slow'' adiabatic temperature change measurements were performed in a home-built device using two nested Halbach magnets with a maximum sinusoidal field change of  $\mu_{0} \Delta H$ of \SI{1.93}{T} with a starting field-sweep rate of \SI{0.93}{Ts^{-1}}. For measurements of the adiabatic temperature change, a differential \textit{T}-type thermocouple was glued between a stack of two sample plates with epoxy of high thermal conductivity. ``Fast'' $\Delta T_{ad}$ measurements were carried out in pulsed magnetic fields up to \SI{20}{T} at the Dresden High Magnetic Field Laboratory (HLD). For all pulse-field measurements shown here, the maximum field is reached within \SI{0.013}{s} with a maximum field-sweep rate of up to \SI{2870}{Ts^{-1}} achieved in a total field change of  \SI{20}{T}. A thin differential \textit{T}-type thermocouple of \SI{25}{\micro m} wire thickness was attached similar to the ``slow'' measurement allowing to detect the temperature change with fast response time. 

In order to correct the temperature measurement with respect to field-induced voltages into the thermocouple, a background signal was recorded at high temperature approximately \SI{50}{K} above the phase transition and subtracted from the measured signals. The absolute sample temperature was determined via a Pt100 thermometer which was located close to the sample. The magnetic field was recorded using a pick-up coil. Previous measurements using this setup showed that eddy-current heating of the samples is negligible~\cite{Gottschall2016-2, Zavareh2015, Scheibel2015}. The same was is true for the present measurements as the resistivity of the Fe$_2$P sample is lower compared to the previously measured Heusler and La-Fe-Si samples. For all measurements of $\Delta T_{ad}$, a non-continuous measurements protocol was chosen~\cite{Gottschall2015}. The initial state of the material was restored after every field-application cycle by heating the sample to \SI{300}{K} i.e. the paramagnetic state and subsequent cooling to the desired starting temperature. This step is necessary in order to avoid the effects of thermal hysteresis inherent to first-order materials as the $\Delta T_{ad}$ would be considerably lower on the second and following field cycles~\cite{Gutfleisch2016}.

\section{Experimental Results and Discussion}

\begin{figure}
\includegraphics[scale=1]{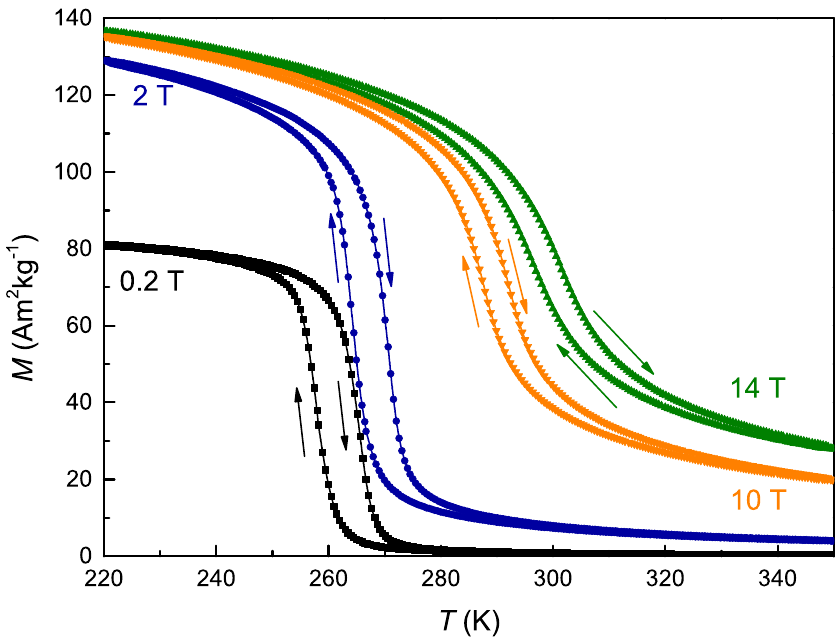} 
\caption{Magnetization versus temperature for fields of 0.2, 2, 10, and \SI{14}{T}.} 
\label{fig:MT-TtvsH}
\end{figure}

Isofield magnetization measurements versus temperature are plotted in Fig.~\ref{fig:MT-TtvsH} for fields of 0.2, 2, 10, and \SI{14}{T}. The ferromagnetic (fm) to paramagnetic (pm) transition starts at \SI{250}{K} and the material is fully transformed at \SI{270}{K} in \SI{0.2}T, accompanied by a thermal hysteresis of about \SI{7}{K} between the heating and cooling branch. In larger magnetic fields the transition shifts to higher temperatures and the transition width broadens drastically with a reduced thermal hysteresis of about \SI{4.5}{K} in \SI{14}{T}. Additionally, the magnetization in the paramagnetic regime is largely increased in comparison to the value measured in \SI{0.2}{T}.

\begin{figure}
\includegraphics[scale=1]{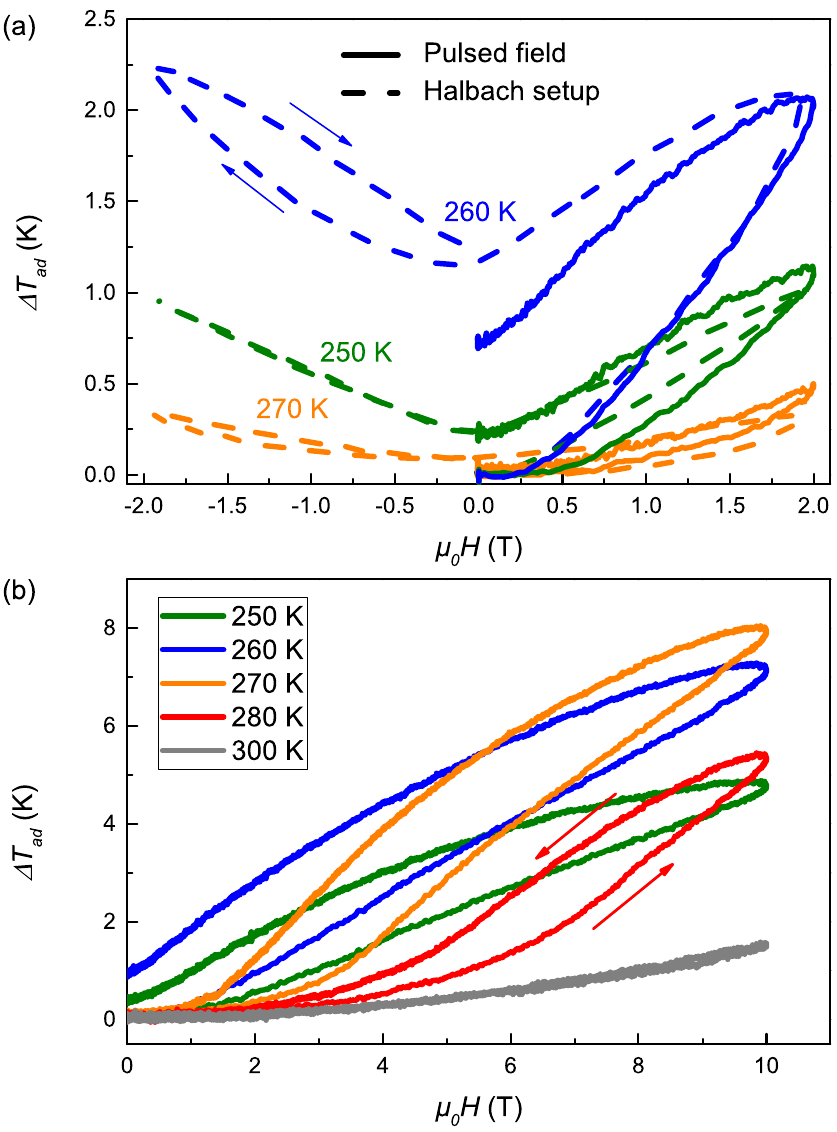} 
\caption{(a) Field-dependent $\Delta T_{ad}$ measured in a Halbach and pulsed-field-magnet setup with a field change of 1.93 and \SI{2}{T}, respectivel,y and in (b) for pulsed fields of \SI{10}{T}.}
\label{fig:DT-2Tvst-darmstadt}
\end{figure}

The field-dependent adiabatic temperature change measured for ``slow'' (\SI{1.93}{T}) and ``fast'' (\SI{2}{T}) field-sweeps is shown in Fig.~\ref{fig:DT-2Tvst-darmstadt}(a) for temperatures above and below the transition. At a temperature of \SI{250}{K} for the ``slow'' field application (dashed curve), an adiabatic temperature change $\Delta T_{ad}$ of \SI{1}{K} is observed in the first application cycle, matching the value in the second field-application cycle to \SI{-1.93}{T}. The material shows a reversible $\Delta T_{ad}$ of only \SI{0.75}{K} due to the thermal hysteresis, inherent to a first-order phase transition.

At \SI{250}{K}, before field application, the material has not yet fully transformed to a pure fm state, but is locked in a mixed state due to the width of the transition. This behavior is not uncommon for first-order-transition materials and has been found e.g. in Heusler alloys \cite{Gottschall2016-3}. A mixed state can result from local chemical inhomogeneity of the sample which was already shown for similar Fe$_2$P-type alloys \cite{Bartok2016}. Also mechanical stress building up in the sample due to the anisotropic lattice expansion during the phase transition~\cite{Dung2011-2} hinders the nucleation and growth process \cite{Gottschall2017}. The second field-application cycle, when the magnetic field is applied in the opposite direction, is fully reversible. Therefore, one can assume that due to the application of the magnetic field in the first cycle, all material is transformed to the pure fm state and a conventional fully reversible magnetocaloric effect appears. 

At \SI{260}{K}, the largest $\Delta T_{ad}$ of \SI{2.1}{K} is observed. An opening of the $\Delta T_{ad}(H)$ loops is visible for the first and second field-application cycle. Here, the material transforms in a minor magnetization loop in a mixed-state region when compared to Fig.~\ref{fig:MT-TtvsH}. The back and forth transformation follows two non-linear field-dependent paths while undergoing the transition resulting in a distinct field-dependent hysteresis.  This kind of complex behavior has been previously discussed by Gottschall $et \ al.$ for Heusler alloys and was ascribed to the nucleation and growth process of the martensite and austenite phase \cite{Gottschall2015}.  At the starting temperature of \SI{270}{K} (above the fm to pm transition), the second loop shows a slight hysteresis indicative for an incipient first-order behavior. 

The field-sweep rate in the pulsed-field setup is much higher than in the permanent-magnet Halbach-array setup, yet only one cycle can be performed per experiment. In the pulsed-field setup, the field of \SI{2}{T} is reached within \SI{0.013}{s} with a maximal field-sweep rate of \SI{235}{Ts^{-1}} and at \SI{260}{K}, the highest temperature change of \SI{2.1}{K} is the same as for the conventional ``slow'' field sweep rates. Both experiments exhibit similar $\Delta T_{ad}$ dependencies whereas the reversible part of $\Delta T_{ad}$ is larger for the pulsed-field setup.

Measurements with a maximum field-sweep rate of  \SI{1330}{Ts^{-1}} performed in pulsed-fields of \SI{10}{T} are shown in Fig.~\ref{fig:DT-2Tvst-darmstadt}(b) . A maximum $\Delta T_{ad}$ of \SI{8}{K} is observed at \SI{270}{K}. At a starting temperature of \SI{250}{K}, the temperature of the sample changes directly with applied magnetic field and shows hysteresis similar as discussed above in Fig.~\ref{fig:DT-2Tvst-darmstadt}(a). This behavior is independent of the field-sweep rate and the field of \SI{10}{T} is not large enough to fully saturate the fm state, as can also be seen in the slight irreversibility of the temperature change. The largest thermal hysteresis is present at a starting temperature of \SI{260}{K}. At \SI{270}{K}, a change in slope can be observed at around \SI{3}{T} for increasing and at \SI{1}{T} for decreasing the field. The same holds true for measurements performed at \SI{280}{K}. For both cases, the samples is cooled back to the initial value, as the starting temperatures are far away from the transition at zero field. Yet, the temperature does not follow the same path while transforming from the pm to fm state and back again. At \SI{300}{K}, a hysteresis-free magnetocaloric effect originating from the partial alignment of magnetic moments in the paramagnetic region is shown as the starting temperature is too high to induce a first-order transition in \SI{10}{T}.

\begin{figure}
\includegraphics[scale=1]{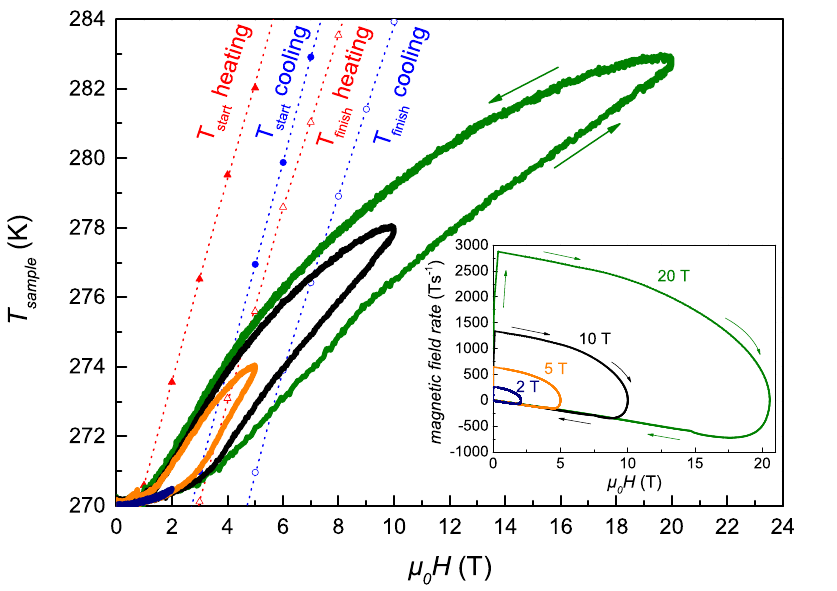} 
\caption{Sample temperature as a function of magnetic field in pulsed field changes of 2, 5, 10, and \SI{20}{T}. The corresponding phase lines for heating and cooling curves for the start and finish temperatures of the phase transition are indicated by dotted lines. The inset shows the field dependence of  the magnetic-field sweep rate for the pulses.}
\label{fig:DT-20Tvst}
\end{figure}

In order to study this behavior in more detail, measurements of $\Delta T_{ad}$ were performed at \SI{270}{K} with field changes of 2, 5, 10, and \SI{20}{T}, shown in Fig~\ref{fig:DT-20Tvst}. All measurements started from the same thermodynamic state as described in the experimental section. All field pulses reached their maximum after \SI{13}{ms} resulting in field-sweep rates of 235 to \SI{2850}{Ts^{-1}}.

In \SI{20}{T}, an adiabatic temperature change of \SI{13}{K} could be measured.
Up to a field of about \SI{3}{T} the curves overlap showing a slight temperature increase attributed to a magnetocaloric effect which is present due to the alignment of magnetic moments in the paramagnetic state. Here, the transition is sweep-rate independent and the material can follow the field changes.  A change in the slope of $\Delta T_{ad}(H)$ above \SI{3}{T} is observed for all field-sweep rates. The change in slope coincides approximately with the starting temperature of the transition, determined by the onset of the magnetization curves shown in Fig.~\ref{fig:MT-TtvsH}. This is indicated by the dashed lines in Fig.~\ref{fig:DT-20Tvst} marking the start and finish temperatures of the first-order field-induced phase transition which do not overlap due to thermal hysteresis resulting in a field hysteresis of the $\Delta T_{ad}(H)$ curves. This is a distinct feature of the transition as it can also be observed in the measurements of $\Delta T_{ad}$ with the Halbach setup, shown in Fig.~\ref{fig:DT-2Tvst-darmstadt}(a). However, the curves measured at different field-sweep rates in pulsed fields show divergent behavior. A bad thermal contact of the sample with the thermocouple or a slow response time as origin is excluded since the up and down field-sweep data in Fig.~\ref{fig:DT-2Tvst-darmstadt}(b) at \SI{300}{K} lie on top of each other.  As the slope of the curves varies with the different pulses we conclude that the first-order phase transition cannot entirely follow the fast field changes and the nucleation and growth of the fm hexagonal phase formation lacks behind. However, no quantitative estimate on the actual dynamics of the transition can be given by these measurement.

The distinct field dependence of $\Delta T_{ad}$ can be explained by the profile of the field pulse as shown in the inset of Fig.~\ref{fig:DT-20Tvst}. As all pulses, independent of the peak field strength, reach their maximum after \SI{13}{ms}, the field-sweep rate varies with the peak field applied. One finds (Fig.~\ref{fig:DT-20Tvst}) that the slopes of the increasing field curves depend on the sweep rate while the slopes for the decreasing fields, where the field change rates are in generally slower, are similar.

 Even in fields up to \SI{20}{T} $\Delta T_{ad}(H)$ does not saturate as was observed in Heusler alloys and antiperovskites~\cite{Gottschall2016-2, Zavareh2015, Scheibel2015}. In contrast to Fe$_2$P alloys, Heusler and antiperovskites are inverse magnetocaloric materials where the magnetic field shifts the transition to lower temperatures, and the driving force of the magnetocaloric effect, the shift of transition temperature $T_t$ with applied magnetic field $H$ ($\frac{dT_t}{dH}$) becomes larger with higher magnetic fields~\cite{Taubel2017a}. 
A non-saturating behavior on the other hand was also observed in Fe$_2$P-type alloys where the isothermal entropy change was studied in fields up to \SI{10}{T} \cite{Fries2017}. In this material class, a continuously growing magnetocaloric effect due to a field-induced partial alignment of magnetic moments in the paramagnetic phase overlaps with the first-order transition, therefore, exhibiting no saturation of $\Delta T_{ad}$. This can be partly attributed to the fact that in conventional materials an applied magnetic field shifts $T_t$ towards higher temperatures. However, the shift decelerates with increasing field strength. For instance, the shift of $T_t$ from 0.2 to \SI{2}{T} is nearly the same as for a field change from 10 to \SI{14}{T}, see Fig.~\ref{fig:MT-TtvsH}. Therefore, the magnetic fields required to complete the transition are higher in conventional materials than for example in inverse magnetocaloric materials.
 
\begin{figure}
\includegraphics[scale=1]{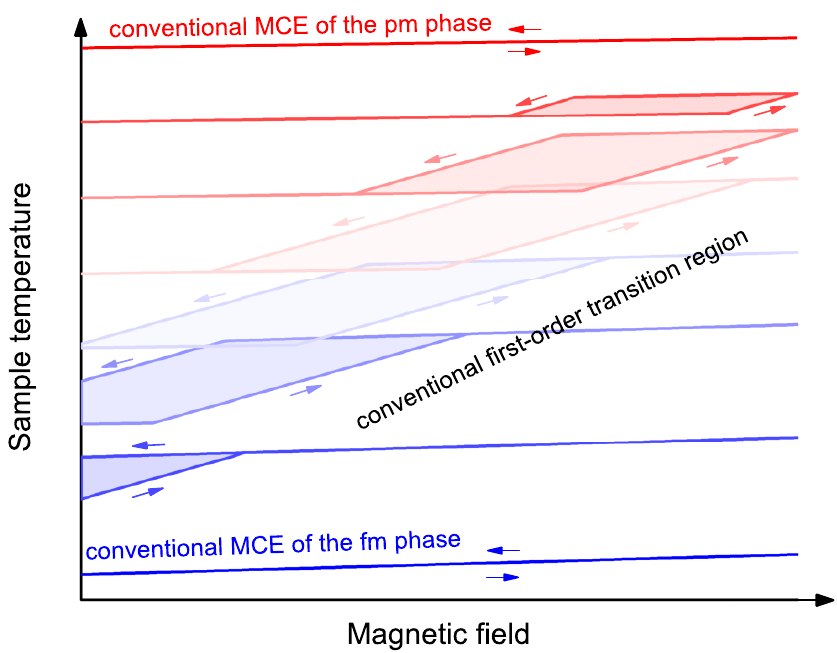} 
\caption{Schematic of a field-dependent temperature change for a typical magnetocaloric material with first-order transition at distinct starting temperatures where a complete transformation is possible.}
\label{fig:completed-trans}
\end{figure} 
 
Considering the shift of transition temperature with field, the first-order phase transition should be completed within the 10 and \SI{20}{T} field changes. A schematic of a field-dependent $\Delta T_{ad}$ change for a first-order transition is shown in Fig.~\ref{fig:completed-trans}.  Three different contributions to the temperature change can be distinguished: First, the alignment of magnetic moments in the ferromagnetic phase contributes to the temperature change (for a starting temperature far below the transition temperature), second the magnetocaloric effect from the alignment of magnetic moments in the paramagnetic regime (shown for the case of a starting temperature far above the transition temperature). The third contribution is the MCE arising from the actual first-order transition which is shown for a complete transition at temperatures close to the transition temperature. All of them are accompanied by a change in slope of the $\Delta T_{ad}(H)$ curve. For Mn-Fe-P-Si, the transition shifts towards higher temperatures in field. Therefore, higher magnetic fields are required to induce the transformation when starting more and more above the transition temperature. In case of a complete transition, the magnetocaloric effects of the ferro- and paramagnetic phase should be hysteresis free and the MCE attributed to the paramagnetic phase should be much smaller compared to the first-order transition. Such a behavior has been observed e.g. for a Heusler alloy~\cite{Gottschall2016-2}.

\begin{figure}
\includegraphics[scale=1]{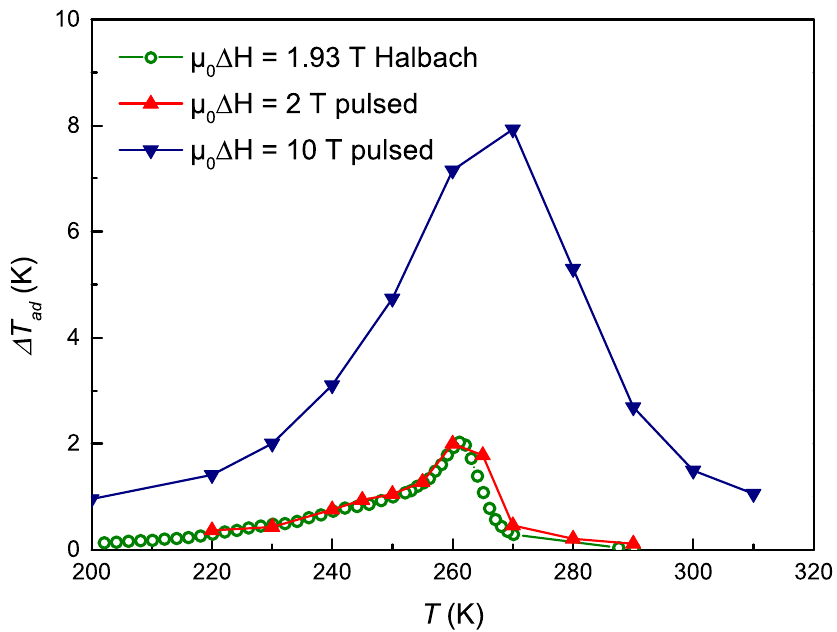} 
\caption{The irreversible adiabatic temperature change versus temperature for discontinuous measurements performed in a Halbach magnet array with a field change of \SI{1.93}{T} and in pulsed fields of 2 and \SI{10}{T}.}
\label{fig:abs-DT-T}
\end{figure}

The maximum values of the  irreversible $\Delta T_{ad}$ are shown in Fig.~\ref{fig:abs-DT-T} for both the ``slow'' and ``fast'' measurements. 
The data measured by use of the Halbach array with a field change of \SI{1.93}{T} and the measurements in pulsed fields of \SI{2}{T} agree very well. This proofs not only the field-rate independence of the maximum $\Delta T_{ad}$ for this field, but also the quality of both measurements.  In the fm region a non-negligible $\Delta T_{ad}$ can be observed which is attributed to a conventional magnetocaloric effect as the magnetization is not saturated in low fields (see Fig.~\ref{fig:MT-TtvsH}). $\Delta T_{ad}$ for \SI{10}{T} pulsed-field measurements shows a similar temperature dependence as the \SI{2}{T} data with a shift of the maximum in $\Delta T_{ad}$ towards higher temperatures due to the field-induced shift of the phase transition. In any case, there is no plateau of the MCE forming as would be expected for saturated first-order transitions (see Fig.~\ref{fig:completed-trans}). This implies that 10, and even \SI{20}{T}, are not yet sufficient to completely transform the material.

\section{Summary}

In summary, $\Delta T_{ad}$ was measured with field-sweep rates ranging from \SI{0.93}{Ts^{-1}} to \SI{2870}{Ts^{-1}}. A saturation of the magnetocaloric effect was not observed in fields up to \SI{20}{T} as the MCE due to the field-induced partial alignment of magnetic moments the paramagnetic regime always overlaps with the first-order transition, and the first-order transition can not be completed. It is shown that the first-order phase transition is field-rate dependent contrary to a previous report \cite{Cugini2016}. However, the maximum values of $\Delta T_{ad}$ seem to be independent of the sweep rate at least up to field-sweep rates of \SI{235}{Ts^{-1}}. The MCE of the pure fm and pm phase can follow all field-sweep rates. For applications it follows that the material is suitable in a real magnetic cooling device with field-change frequencies below \SI{100}{Hz}. Although a frequency-dependent increase the hysteresis would lower the efficiency of the material.
\appendix

\section*{Acknowledgments}

The work was supported from the European Research Council (ERC) under the European Unions Horizon 2020 research and innovation programme (grant agreement No 743116 - project Cool Innov).
We thank the DFG (Grant No. SPP 1599) and the European Community’s Seventh Framework Programme FP7/2007-2013 under grant agreement number 310748 (DRREAM) for financial support.  The support from HLD at HZDR, member of the European Magnetic Field Laboratory (EMFL) is acknowledged. We also thank L. V. B. Diop for experimental advice.

\section*{References}

\bibliography{mylib}

\begin{thebibliography}{10}
\expandafter\ifx\csname url\endcsname\relax
  \def\url#1{\texttt{#1}}\fi
\expandafter\ifx\csname urlprefix\endcsname\relax\def\urlprefix{URL }\fi
\expandafter\ifx\csname href\endcsname\relax
  \def\href#1#2{#2} \def\path#1{#1}\fi

\bibitem{Morna2009}
M.~Isaac, D.~P. van Vuuren, Modeling global residential sector energy demand
  for heating and air conditioning in the context of climate change, Energy
  Policy 37~(2) (2009) 507--521.
\newblock \href
  {http://dx.doi.org/http://dx.doi.org/10.1016/j.enpol.2008.09.051}
  {\path{doi:http://dx.doi.org/10.1016/j.enpol.2008.09.051}}.

\bibitem{Faehler2012}
S.~F\"ahler, U.~K. R\"o{\ss}ler, O.~Kastner, J.~Eckert, G.~Eggeler,
  H.~Emmerich, P.~Entel, S.~M\"uller, E.~Quandt, K.~Albe, {Caloric Effects in
  Ferroic Materials: New Concepts for Cooling}, Adv. Eng. Mater. 14~(1-2)
  (2012) 10--19.
\newblock \href {http://dx.doi.org/10.1002/adem.201100178}
  {\path{doi:10.1002/adem.201100178}}.

\bibitem{Gutfleisch2011}
O.~Gutfleisch, M.~A. Willard, E.~Br\"uck, C.~H. Chen, S.~G. Sankar, J.~P. Liu,
  {Magnetic Materials and Devices for the 21st Century: Stronger, Lighter, and
  More Energy Efficient}, Adv. Mater. 23~(7) (2011) 821--842.
\newblock \href {http://dx.doi.org/10.1002/adma.201002180}
  {\path{doi:10.1002/adma.201002180}}.

\bibitem{Moya2014}
X.~Moya, S.~Kar-Narayan, N.~D. Mathur, Caloric materials near ferroic phase
  transitions, Nat Mater 13~(5) (2014) 439--450.
\newblock \href {http://dx.doi.org/10.1038/nmat3951}
  {\path{doi:10.1038/nmat3951}}.

\bibitem{Gottschall2018a}
T.~Gottschall, A.~Gràcia-Condal, M.~Fries, A.~Taubel, L.~Pfeuffer, L.~Mañosa,
  A.~Planes, K.~P. Skokov, O.~Gutfleisch, A multicaloric cooling cycle that
  exploits thermal hysteresis, Nature Materials\href
  {http://dx.doi.org/10.1038/s41563-018-0166-6}
  {\path{doi:10.1038/s41563-018-0166-6}}.

\bibitem{Kitanovski2014-book}
A.~Kitanovski, J.~Tu{\u{s}}ek, U.~Tomc, U.~Plaznik, M.~O{\u{z}}bolt,
  A.~Poredo{\u{s}}, {Magnetocaloric Energy Conversion: From Theory to
  Applications}, Springer International Publishing, 2014.

\bibitem{Weiss1917}
P.~Weiss, A.~Piccard, Le ph\'{e}nom\`{e}ne magn\'{e}tocalorique, J. Phys.
  (Paris) 7~(1) (1917) 103--109.
\newblock \href {http://dx.doi.org/10.1051/jphystap:019170070010300}
  {\path{doi:10.1051/jphystap:019170070010300}}.

\bibitem{Brown1976}
G.~V. Brown, Magnetic heat pumping near room temperature, J. Appl. Phys. 47~(8)
  (1976) 3673--3680.
\newblock \href {http://dx.doi.org/10.1063/1.323176}
  {\path{doi:10.1063/1.323176}}.

\bibitem{Kuzmin2007}
M.~D. Kuz’min, Factors limiting the operation frequency of magnetic
  refrigerators, Appl. Phys. Lett. 90~(25) (2007) 251916.
\newblock \href {http://dx.doi.org/10.1063/1.2750540}
  {\path{doi:10.1063/1.2750540}}.

\bibitem{Smith2012}
A.~Smith, C.~R. Bahl, R.~Bj{\o{}}rk, K.~Engelbrecht, K.~K. Nielsen, N.~Pryds,
  {Materials Challenges for High Performance Magnetocaloric Refrigeration
  Devices}, Adv. Energy Mater. 2~(11) (2012) 1288--1318.
\newblock \href {http://dx.doi.org/10.1002/aenm.201200167}
  {\path{doi:10.1002/aenm.201200167}}.

\bibitem{Liu2012}
J.~Liu, T.~Gottschall, K.~P. Skokov, J.~D. Moore, O.~Gutfleisch, Giant
  magnetocaloric effect driven by structural transitions, Nat. Mater. 11~(7)
  (2012) 620--626.
\newblock \href {http://dx.doi.org/10.1038/nmat3334}
  {\path{doi:10.1038/nmat3334}}.

\bibitem{Gutfleisch2016}
O.~Gutfleisch, T.~Gottschall, M.~Fries, D.~Benke, I.~Radulov, K.~P. Skokov,
  H.~Wende, M.~Gruner, M.~Acet, P.~Entel, M.~Farle, Mastering hysteresis in
  magnetocaloric materials, Phil. Trans. R. Soc. A 374 (2016) 20150308.
\newblock \href {http://dx.doi.org/10.1098/rsta.2015.0308}
  {\path{doi:10.1098/rsta.2015.0308}}.

\bibitem{Scheibel2018a}
F.~Scheibel, T.~Gottschall, A.~Taubel, M.~Fries, K.~P. Skokov, A.~Terwey,
  W.~Keune, K.~Ollefs, H.~Wende, M.~Farle, M.~Acet, O.~Gutfleisch, M.~E.
  Gruner, Hysteresis design of magnetocaloric materials—from basic mechanisms
  to applications, Energy Technology 6~(8) (2018) 1397--1428.
\newblock \href {http://dx.doi.org/10.1002/ente.201800264}
  {\path{doi:10.1002/ente.201800264}}.

\bibitem{Franco2012}
V.~Franco, J.~S. Blazquez, B.~Ingale, A.~Conde, {The Magnetocaloric Effect and
  Magnetic Refrigeration Near Room Temperature: Materials and Models}, Annu.
  Rev. Mater. Res. 42 (2012) 305--342.
\newblock \href {http://dx.doi.org/10.1146/annurev-matsci-062910-100356}
  {\path{doi:10.1146/annurev-matsci-062910-100356}}.

\bibitem{Yan2006}
A.~Yan, K.~H. M\"uller, L.~Schultz, O.~Gutfleisch, {Magnetic entropy change in
  melt-spun MnFePGe (invited)}, J. Appl. Phys. 99~(8) (2006) 08K903--4.
\newblock \href {http://dx.doi.org/10.1063/1.2162807}
  {\path{doi:10.1063/1.2162807}}.

\bibitem{Yibole2014}
H.~Yibole, F.~Guillou, L.~Zhang, N.~H.~v. Dijk, E.~Br\"uck, {Direct measurement
  of the magnetocaloric effect in MnFe(P, X )( X = As, Ge, Si) materials}, J.
  Phys. D: Appl. Phys. 47~(7) (2014) 075002.
\newblock \href {http://dx.doi.org/10.1088/0022-3727/47/7/075002}
  {\path{doi:10.1088/0022-3727/47/7/075002}}.

\bibitem{Fries2017}
M.~Fries, L.~Pfeuffer, E.~Bruder, T.~Gottschall, S.~Ener, L.~V.~B. Diop,
  T.~Gr\"ob, K.~P. Skokov, O.~Gutfleisch, {Microstructural and magnetic
  properties of Mn-Fe-P-Si (Fe$_2$P-type) magnetocaloric compounds}, Acta
  Mater. 132 (2017) 222--229.
\newblock \href {http://dx.doi.org/10.1016/j.actamat.2017.04.040}
  {\path{doi:10.1016/j.actamat.2017.04.040}}.

\bibitem{Gauss2016}
R.~Gau\ss, G.~Homm, O.~Gutfleisch, {The Resource Basis of Magnetic
  Refrigeration}, J. Ind. Ecol.\href {http://dx.doi.org/10.1111/jiec.12488}
  {\path{doi:10.1111/jiec.12488}}.

\bibitem{Guillou2014-2}
F.~Guillou, G.~Porcari, H.~Yibole, N.~van Dijk, E.~Br\"uck, {Taming the
  First-Order Transition in Giant Magnetocaloric Materials}, Adv. Mater.
  26~(17) (2014) 2671--2675.
\newblock \href {http://dx.doi.org/10.1002/adma.201304788}
  {\path{doi:10.1002/adma.201304788}}.

\bibitem{Guillou2014-3}
F.~Guillou, H.~Yibole, N.~H. van Dijk, L.~Zhang, V.~Hardy, E.~Br\"uck, {About
  the mechanical stability of MnFe(P,Si,B) giant-magnetocaloric materials}, J.
  Alloys Compd. 617 (2014) 569--574.
\newblock \href {http://dx.doi.org/10.1016/j.jallcom.2014.08.061}
  {\path{doi:10.1016/j.jallcom.2014.08.061}}.

\bibitem{Dung2011-2}
N.~H. Dung, Z.~Q. Ou, L.~Caron, L.~Zhang, D.~T.~C. Thanh, G.~A. de~Wijs, R.~A.
  de~Groot, K.~H.~J. Buschow, E.~Br\"uck, {Mixed Magnetism for Refrigeration
  and Energy Conversion}, Adv. Energy Mater. 1~(6) (2011) 1215--1219.
\newblock \href {http://dx.doi.org/10.1002/aenm.201100252}
  {\path{doi:10.1002/aenm.201100252}}.

\bibitem{Cugini2016}
F.~Cugini, G.~Porcari, C.~Viappiani, L.~Caron, A.~O. dos Santos, L.~P. Cardoso,
  E.~C. Passamani, J.~R.~C. Proveti, S.~Gama, E.~Br\"uck, M.~Solzi,
  {Millisecond direct measurement of the magnetocaloric effect of a
  Fe$_2$P-based compound by the mirage effect}, Appl. Phys. Lett. 108~(1)
  (2016) 012407.
\newblock \href {http://dx.doi.org/10.1063/1.4939451}
  {\path{doi:10.1063/1.4939451}}.

\bibitem{Gottschall2016-2}
T.~Gottschall, K.~P. Skokov, F.~Scheibel, M.~Acet, M.~G. Zavareh, Y.~Skourski,
  J.~Wosnitza, M.~Farle, O.~Gutfleisch, {Dynamical Effects of the Martensitic
  Transition in Magnetocaloric Heusler Alloys from Direct $\Delta T_{ad}$
  Measurements under Different Magnetic-Field-Sweep Rates}, Phys. Rev. Appl
  5~(2) (2016) 024013.

\bibitem{Zavareh2015}
M.~G. Zavareh, C.~S. Mejía, A.~K. Nayak, Y.~Skourski, J.~Wosnitza, C.~Felser,
  M.~Nicklas, {Direct measurements of the magnetocaloric effect in pulsed
  magnetic fields: The example of the Heusler alloy
  Ni$_{50}$Mn$_{35}$In$_{15}$}, Appl. Phys. Lett. 106~(7) (2015) 071904.
\newblock \href {http://dx.doi.org/10.1063/1.4913446}
  {\path{doi:10.1063/1.4913446}}.

\bibitem{Scheibel2015}
F.~Scheibel, T.~Gottschall, K.~Skokov, O.~Gutfleisch, M.~Ghorbani-Zavareh,
  Y.~Skourski, J.~Wosnitza, O.~Cakir, M.~Farle, M.~Acet, {Dependence of the
  inverse magnetocaloric effect on the field-change rate in Mn$_3$GaC and its
  relationship to the kinetics of the phase transition}, J. Appl. Phys.
  117~(23) (2015) 233902.
\newblock \href {http://dx.doi.org/10.1063/1.4922722}
  {\path{doi:10.1063/1.4922722}}.

\bibitem{Gottschall2015}
T.~Gottschall, K.~P. Skokov, B.~Frincu, O.~Gutfleisch, {Large reversible
  magnetocaloric effect in Ni-Mn-In-Co}, Appl. Phys. Lett. 106~(2) (2015)
  021901.
\newblock \href {http://dx.doi.org/10.1063/1.4905371}
  {\path{doi:10.1063/1.4905371}}.

\bibitem{Gottschall2016-3}
T.~Gottschall, K.~P. Skokov, R.~Burriel, O.~Gutfleisch, {On the S(T) diagram of
  magnetocaloric materials with first-order transition: Kinetic and cyclic
  effects of Heusler alloy}s, Acta Mater. 107 (2016) 1--8.
\newblock \href {http://dx.doi.org/10.1016/j.actamat.2016.01.052}
  {\path{doi:10.1016/j.actamat.2016.01.052}}.

\bibitem{Bartok2016}
A.~Bartok, M.~Kustov, L.~F. Cohen, A.~Pasko, K.~Zehani, L.~Bessais,
  F.~Mazaleyrat, M.~LoBue, {Study of the first paramagnetic to ferromagnetic
  transition in as prepared samples of Mn–Fe–P–Si magnetocaloric
  compounds prepared by different synthesis routes}, J. Magn. Magn. Mater. 400
  (2016) 333--338.
\newblock \href {http://dx.doi.org/10.1016/j.jmmm.2015.08.045}
  {\path{doi:10.1016/j.jmmm.2015.08.045}}.

\bibitem{Gottschall2017}
T.~Gottschall, D.~Benke, M.~Fries, A.~Taubel, I.~A. Radulov, K.~P. Skokov,
  O.~Gutfleisch, {A matter of size and stress: Understanding the first-order
  transition of materials for solid-state refrigeration}, Adv. Funct. Mater.
  27~(32) (2017) 1606735.
\newblock \href {http://dx.doi.org/10.1002/adfm.20160673}
  {\path{doi:10.1002/adfm.20160673}}.

\bibitem{Taubel2017a}
A.~Taubel, T.~Gottschall, M.~Fries, S.~Riegg, C.~Soon, K.~P. Skokov,
  O.~Gutfleisch, A comparative study on the magnetocaloric properties of
  ni-mn-x(-co) heusler alloys, physica status solidi (b) 255~(2) (2017)
  1700331.
\newblock \href {http://dx.doi.org/10.1002/pssb.201700331}
  {\path{doi:10.1002/pssb.201700331}}.

\end{thebibliography}

\end{document}